Bringing Order to the Chaos in the Brickyard

Bethany Shifflett

San Jose State University

**Abstract**

An allegory published in 1963 titled 'Chaos in the Brickyard' spoke to the decline in the quality of research. In the intervening time greater awareness of the issues and actions to improve research endeavors have emerged. Still, problems persist. This paper is intended to clarify some of the challenges, particularly with respect to quantitative research, then suggest ways to improve the quality of published research. The paper highlights where feasible refinements in analytical techniques can be made and provides a guide to fundamental principles related to data analysis in research.

**Introduction**

Forscher's (1963) allegory portrayed scientists as builders constructing edifices (theory) by assembling bricks (facts). As the story 'Chaos in the Brickyard' explains, the original pride in producing bricks of the highest quality so as to facilitate the creation of solid edifices gave way to an obsession with simply making bricks. 'Unfortunately, the builders were almost destroyed. It became difficult to find the proper bricks for a task because one had to hunt among so many' (Forscher, 1963, p. 339). The ripple effect of this piece can be observed through faculty who continue to introduce their students to this commentary in order to elicit an awareness of significant design and analysis issues in research. Faculty members (from wide-ranging disciplines) may take the discussion in a particular direction (e.g., ethics, data integrity, reporting bias); yet the overall impact has likely been that students become more familiar with the problematic nature of published work than they would have been. In fact, in the author's experience, students often realize for the first time that published work might be flawed after reading 'Chaos in the Brickyard'. Forscher's story provides a springboard for faculty to continue the dialogue with their students and an opportunity for researchers to reflect on the status of published research today.

Turning attention then to the identification of problems along with possible strategies to address areas where publications fall short can help guide efforts to diminish the chaos in the brickyard. Hence, this paper seeks to focus on the foundational elements of quantitative research in order to emphasize the importance of basic principles pertaining to research design, descriptive statistics, and inferential statistics. The intention is to pull together from an array of texts and articles one accessible resource for researchers to consider and for faculty to integrate with other instructional materials as they continue the discussion begun by Forscher.

**Challenges**

Awareness of the wide range of issues related to the quality of publications has certainly been raised through academic journal publications (Fischman, 2011; Knudson, 2009; Weed, 2006) and the media in general (Bower, 2013; Chwe, 2014; Kolata, 2013; Lamb, 2013; Weins, 2014). Researchers note that what continues to be a problem, and perhaps a more pronounced one since Forscher's time, is the proliferation of research of questionable quality (Bauerlein, Gad-el-Hak, Grody, McKelvey & Trimble, 2010). It is important to make a distinction at this point between the proliferation of weak research and the proliferation of data. The problem is not the explosion of available data, often referred to as 'big data'. Big data are here to stay and researchers are beginning to understand how best to capture and use it to good effect. A particularly good example is the work done by Silver (2012) in analyzing large volumes of data to predict election results. His success leant credibility and respect to an analytical approach to practical problem solving. The visibility of this quantitative work provides an opportunity to garner support and build public confidence in research provided publications possess comparable credibility and quality.

Consider the point made by Baerlein et al., (2010) that the 'amount of redundant, inconsequential, and outright poor research has swelled in recent decades' (p. 1). Taken as a call to change the landscape, it is a challenge worth tackling. The situation has a particularly negative impact on everyone involved when one considers the time required by researchers to read and evaluate volumes of published work to determine its relevance, quality, and connection to their own projects in addition to the time invested in having the work assessed by editors and reviewers. A related problem is the proliferation of online journals that appear to publish work without genuine peer review or consideration of the quality of the research provided a fee is paid by the author (Beall, 2013; Kolata, 2013). The expansion of new open-access online publication venues does increase the opportunity for the expedient distribution of research and arXiv.com is an example of how valuable credible online resources can be. The challenge for the researcher becomes identifying reputable online journals from among so many. A confluence of ongoing pressures on faculty to publish combined with the predatory nature of a growing number of publishers of questionable integrity may be exacerbating the 'chaos in the brickyard'. When the field becomes littered with poor quality research the task of simply finding solid work to build on becomes a challenge.

**Issues: Bias and Fragmentation**

Among the issues that frequently receive attention is bias. For example, selection bias is the practice, often associated with government agencies, businesses, and the pharmaceutical industry, of being selective in the reporting of research/evidence to the point where findings are misrepresented. Reporting bias, on the other hand, is the predisposition to give less attention to, choose not to submit for review, or not publish work with 'negative' results (Editorial commentary, 2007; Pigott, Valentine, Polanin, Williams, & Canada, 2013). Such bias could lead to conclusions that treatments are more useful than otherwise might be the case if research with both significant and non-significant findings were viewed as relevant. One indication of the problematic situation is the observation by Ioannidis (2005) that data mining resources are publicized for their capacity to identify significant results. This puts at the top of the list of priorities finding something (anything) significant rather than identifying and exploring important and relevant questions. One additional issue in this category is confirmation bias. This pertains to giving less scrutiny to results in line with expectations. Picture the deep and probing review of data entry, error checking, and appraisal of analytical procedures that might ensue when findings of a completely unexpected nature occur. Does that same level of scrutiny take place when findings are in line with expectations? If not, the likelihood that confirmation bias may lead to the perpetuation of inaccurate findings is cause for concern.

Another problematic practice is the piecemeal or fragmented publication of research findings. Referred to by Fischman (2011) as 'salami science', the practice of publishing multiple articles all derived from one study can misrepresent the extent to which findings are statistically significant. It also gives the illusion of greater breadth and depth of study in an area than has actually taken place. While researchers have recommended the use of meta-analysis to better assemble all the various studies in an area (Altman, 2012; Knudson, 2009; Weed, 2006; Weed, 2005) the original problem of having it appear that multiple independent studies have been conducted remains.

**Issues: Research Methods and Data Analysis**

Two of the issues raised in Forscher's story are equally important. The first was the lament that few aspired to be builders. The second was that the poor quality of numerous bricks would inhibit progress. Theory development and testing that emerges from theory-driven questions designed to extend the knowledge base are important (Achterberg, Novak, & Gillespie, 1985; Eisenhardt, 1989; Walshe, 2007) and builders that take us in this direction are needed. Equally important in Forscher's story, and relevant today, is the need for bricks of the highest quality. Some of those bricks will not necessarily be theory-based yet they probe important questions that need exploration. It takes both builders and brick makers to advance our understanding in any discipline. This section focuses on the elements of basic research that impact the quality of the research/bricks produced which can subsequently facilitate, or inhibit if of poor quality, the work of theory building.

With respect to methodological and analytical issues, it is fairly common to find in the critique of research concerns regarding how statistics are used (Bartlett, 2013; Franks & Huck, 1986; Knudson, 2009; Marteniuk & Bertram, 1999; Seife, 2011; Seife, 2014; Taleb, 2014; Vaisrub, 1991). Building quantitative research skills in the process of honing scientific literacy could prove valuable in resolving some of the problems associated with the application of statistics to analyze research data. This does not mean that students, faculty, reviewers, and researchers in all disciplines need to become expert level statisticians. A solid grasp of the basics with regard to descriptive and inferential statistics can, in a substantive way, help bring order to the 'chaos in the brickyard'. For example, central to the selection of descriptive and inferential statistics to summarize group data is an understanding of how the type of data collected impacts what statistics should be used. The information presented in Table 1 and Table 2 provides a basic guide regarding which descriptive and inferential statistics to use depending on the type of data available. For descriptive statistics, discrete data (categorical or ordinal) are best summarized with frequencies or percentages. When continuous data (interval, ratio) are recorded, measures of central tendency (e.g., mean, median) and variability are appropriate for summarizing the data descriptively.

Table 1

*Descriptive Statistics for Summary of Group Data*

| Type of Data | Descriptive statistics |
|---|---|
| Categorical | Frequencies, Percentages, Mode |
| Ordinal | Percentages, Mode, Median* |
| Interval | Median, Mean, Standard Deviation |
| Ratio | Median, Mean, Standard Deviation |

*Note: The median, as a measure of central tendency, for data at the ordinal level of measurement is acceptable provided the data do not simply represent a few ordered categories.

Table 2

*Inferential Statistics for Testing Differences and Relationships (Correlation)*

___________________________________________________________________________

| Type of Data | Type of Question | Inferential Statistics |
|---|---|---|
| Categorical | Differences | Not Applicable |
| | Relationships | Chi Squared |
| Ordinal | Differences | Mann Whitney, Kruskal Wallis, Wilcoxon, Friedman |
| | Relationships | Chi Squared |
| Interval | Differences | t-tests, F tests |
| | Relationships | Correlation, Regression |
| Ratio | Differences | t-tests, F tests |
| | Relationships | Correlation, Regression |

___________________________________________________________________________

When testing hypotheses where the data for the dependent variable are discrete, nonparametric inferential statistics should be employed. When the data are continuous, parametric inferential statistics would be employed provided distributional assumptions are met. The chart in Figure 1 could help researchers in their selection of analytical techniques, and help reviewers identify errors. One additional point is that when discussing the results pertaining to relationships questions, care needs to be taken so that the presence of a correlation is not used to imply causality.

*Figure 1:* Flow chart for the selection of appropriate inferential analyses depending on the type of question and the data's level of measurement.

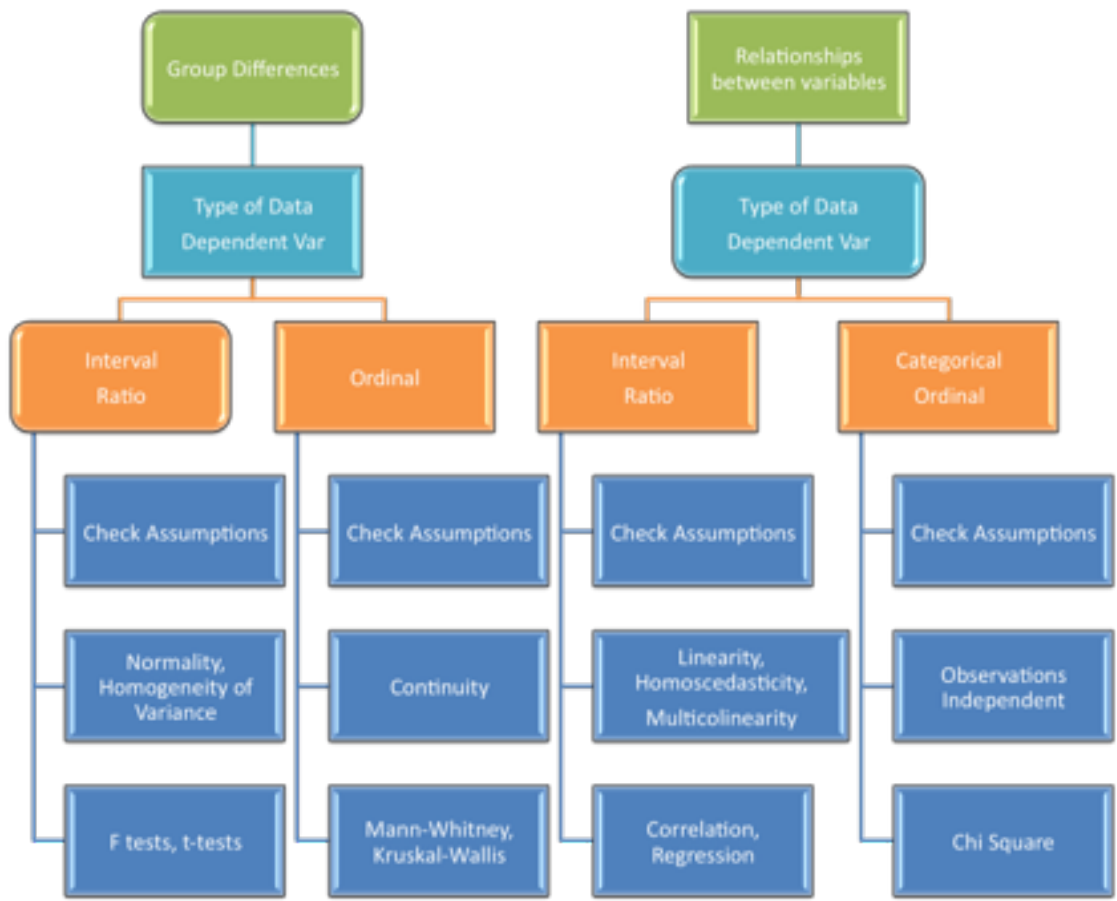

Another concern pertains to the use of demographic data as dependent variables (e.g., age) for inferential tests. Demographic information is best summarized using descriptive rather than inferential statistics. Subsequently, demographic information may be employed as independent variables in an inferential test related to the primary question(s). The distinction pertains to using descriptive statistics to inform subsequent inferential tests rather than conducting inferential tests using demographic characteristics of subjects as dependent variables.

Other basics where a firm understanding is important include (a) the interconnectedness of power, sample size, effect size, and type I error, (b) the importance of reporting effect sizes, p values, and practical significance, (c) the need to adjust the type I error rate prior to examining statistical significance when multiple inferential tests are done, (d) checking distributional assumptions and using non-parametric tests when appropriate, and (e) examining the validity and reliability of data for the dependent variable in a study. The following sections elaborate on these issues.

**Power, Sample size, Effect size, and Type I Error**. Power pertains to the probability of correctly rejecting a null hypothesis and is influenced by sample size, effect size, and selection of alpha (type I error). Effect size conveys the magnitude of the difference or relationship found in a study and type I error is a value selected by the researcher that sets the limit on the probability of incorrectly rejecting a null hypothesis. The important point with regard to the interconnectedness of power and other research design factors is to use the information to determine the sample size needed before a study begins (Myers, Ahn, & Jin, 2011). Once the experiment-wise alpha (type I error), power desired (commonly .80), and expected effect size (identified through pilot studies or previous research) are selected, software can be used to determine sample size (Faul, Erdfelder, Lang, & Buchner, 2007).

**Practical Significance**. Regarding analyses connected to the main question under study, it is important to report the practical significance of the findings. It is all too common for only statistical significance to be reported. Differences or relationships that may be statistically significant are not necessarily of practical or clinical importance (Ioannidis, 2005). For example, finding a statistically significant relationship simply means that you have rejected the null hypothesis that the correlation is zero. Having established it is not zero is not the same as having established that the relationship is of practical importance. With enough subjects, a correlation coefficient of .20 (very weak) could be statistically significant. The more important point, in this example, is that practical significance, reflected in the coefficient of determination (r squared) is only .04 in this case. This tells you that only four percent of the variability in the dependent variable can be explained by the independent variable.

Another point to keep in mind is that what constitutes practical significance can depend on the context. For example, if a measure of leg strength when correlated with vertical jump is .60, then the value reflecting practical significance (r squared = .36) means that 36% of the variability in vertical jump can be explained by leg strength. While this number, absent context, may be interpreted as only modest practical significance, in the context of the fact that other variables like height and muscle fiber type cannot be changed, this value is more likely of substantial practical significance to the researcher, coach, or athlete considering resistance training to improve vertical jump.

The practical relevance of findings is central to theory building and the advancement of ideas. In addition, reporting effect sizes provides scholars with a way to gauge the magnitude of a difference or relationship as well as giving future scholars a key component for meta analyses.

**Adjusting Alpha**.  The need to adjust alpha (type I error) prior to examining statistical significance when multiple tests are done is important. Too often statistical tests are conducted in comparison to an alpha of typically .05 regardless of how many inferential tests are conducted (Franks & Huck, 1986; Ioannidis, 2005; Knudson, Morrow & Thomas, 2014). When the experiment-wise alpha is not adjusted prior to making decisions with respect to statistical significance this could result in exaggerated claims with regard to the significance of findings and caution is called for on the part of readers when this situation is encountered. It is advisable to limit the number of inferential tests conducted. Otherwise it becomes increasingly difficult to obtain statistical significance and power is sacrificed.

When multiple inferential tests are deemed important, the Bonferroni technique (Franks & Huck, 1986) is a simple approach to adjusting alpha and is easily applied; divide the experiment-wise alpha by the number of statistical tests done. This new adjusted alpha is what each p value (probability that the finding is due to chance) is then compared to in determining statistical significance. In addition, when reporting statistical significance it is better to report the actual p value rather than the common 'p < .05' statement (Tromovitch, 2012). It is more informative to the reader and easily obtained from software (e.g., SPSS).

In the author's view, it is not necessary for all questions of interest to be addressed with inferential tests, particularly at the cost of diminished power. The suggestion here is to identify the primary question(s) of interest and apply inferential tests in only those cases. In fact, where feasible, having just one main question addressed inferentially and all other questions handled with descriptive statistics permits the researcher to put all statistical power into that one inferential test. Other questions can be effectively explored via descriptive statistics.

**Distributional Assumptions**. Turning to the topic of checking assumptions, it is an important consideration in the selection of appropriate analyses. For example, if the distributional assumptions associated with the parametric F test from a one-way analysis of variance (ANOVA) are violated, the non-parametric equivalent (Kruskal-Wallis Test) could be used. On the one hand a case can be made that the parametric test is fairly robust to violations of the assumption of normally distributed data and is more powerful than its non-parametric equivalent. On the other hand, if assumptions have been violated the parametric test may misrepresent findings and a comparison based on medians, through a non-parametric test, rather than means may be more appropriate (Thomas, Nelson, & Thomas, 1999). Reporting the outcome after checking assumptions, regardless of the inferential test conducted, serves two valuable functions. First, the reader has been given better context for reported results and second, the need to check assumptions associated with any inferential test gets reinforced and likely replicated by others.

**Psychometrics**. Psychometrics is the last of the topics selected for emphasis. The term psychometrics refers to validity and reliability and can be applied to both research and data. With regard to the validity of research, examining internal and external validity are the key elements. Internal validity is a matter of considering the extent to which findings can be attributed to the independent variable while external validity is about the generalizability of the findings. The reliability of research is focused on the replicability of findings. Good practice calls on the researcher to consider what the threats to internal validity, external validity, and reliability of the research might be and establish protocols for data collection that minimize the threats (Thomas, Nelson, & Silverman, 2010). On this point, the methods section of most publications provides sufficient detail for readers to assess the quality of the research. More problematic is lack of attention to reporting the psychometric characteristics of the data collected.

In examining the reliability of data, of interest is its accuracy. This is typically demonstrated via consistency across repeated measures on one day (internal consistency) or over time (stability). The importance of checking and reporting reliability information for the dependent variable(s) in a study cannot be overstated (Vacha-haase, Ness, Nilsson, & Reetz, 1999). The credibility of all analyses conducted rests on an assumption that the data are accurate. The statistic needed to assess reliability is an intraclass coefficient (e.g., intraclass R or Cronbach's alpha). Though still frequently seen in publications, an interclass coefficient such as the Pearson Product Moment correlation (PPMC) is not the most appropriate statistic for estimating reliability. The PPMC is a rank order correlation coefficient designed to show the relationship between two different variables. It is not designed to detect inconsistency across repeated measures of the same variable; yet consistency is the central issue with reliability.

Regarding evidence of the validity of data, of interest is whether or not the data are clean (not confounded by other factors) and relevant in the context of the research question. Correlating the data from the dependent variable with a criterion measure of the same variable is an appropriate quantitative approach to gathering evidence of the validity of data. Under conditions where a quantitative approach is not feasible (e.g., lack of a criterion measure) at least content validity (cognitive measures) or logical validity (motor skills) should be established through peer review of data collection protocols.

If the quality of the data collected for the dependent variable is questionable then there is little value in testing any hypotheses or trying to draw conclusions from the data. For each study conducted the reliability and validity of the data collected should be examined.

**Simple is elegant.** Vaisrub's (1991) challenge to 'simplify, simplify, in a statistical walden, I dare you' is worth taking note of. With software to handle intricate and otherwise time consuming computations it has become far too easy to point and click through overly complex statistical analyses. When a simple F test from a one-way analysis of variance (ANOVA) will answer the main question in a study, that is what should be done. Conducting instead an excessively complex analysis because one can, contributes directly to the 'chaos in the brickyard'. Even the choice to conduct a 2-way ANOVA should be made in the context of what is needed to answer the primary research question(s) since three F tests will be produced by the 2-way ANOVA. This results in the need to apply a smaller alpha which will make it more difficult to reject null hypotheses and in the process reduce power.

In summary, each of the fundamental design and analysis factors considered in this section, when appropriately incorporated into research increases the likelihood that credible findings will populate the brickyard. Additionally, published work serves as a model for others. This means that each publication with appropriate analyses employed has the potential to influence the quality of research subsequently conducted.

**Recommendations**

Efforts to clear away the chaos generated by poor quality research have been made through critical reviews of published work (Bartlett, 2013; Marteniuk & Bertram, 1999), compilation of an 'authors beware' list of predatory journals (Beall, 2013), and diligent attention among many collegiate faculty to the development of students' scientific and quantitative literacy skills. These efforts to improve the quality of research and publications are important and should continue. In addition, a more widespread approach is recommended for significant and sustainable improvement to help reduce the 'chaos in the brickyard'.

The strategy proposed is one that could be applied to any large-scale project. Imagine a group of colleagues responsible for reviewing the accreditation report and all supporting documentation for their institution. Asking all members to review everything is likely to result in duplication of effort while various components may be overlooked. Alternatively, having each person review a specific portion of the work is more likely to result in a thorough examination of all components. A similar approach can be applied to improve the quality of published research. Students, faculty, and administrators, along with journal editors can each make meaningful contributions and do their part to help bring order to the 'chaos in the brickyard'.

**Student Contributions**. Students are in a position to acquire the skills and knowledge needed to be critical consumers of research and to apply what they learn when the opportunity to conduct research presents itself. With a good understanding of the basics regarding descriptive and inferential statistics they will be better equipped to read and critique publications. In addition, they will be better able to conduct the analyses for projects, theses, and dissertations themselves. While many may not consider quantitative work to be their strong suit, all have the capacity to master the basics to the point where they can think critically about what they read and take responsibility for the design and analysis decisions made for their own research. Faculty can present the information but it is the students who ultimately decide whether or not to embrace the content or simply get through it with no intention to apply what they learn in the future.

**Faculty Contributions**. There are at least three distinct areas where faculty can make a difference: teaching, service and research. At both the undergraduate and graduate levels, attention while teaching directed toward raising the awareness of students with respect to what constitutes quality research can lay a good foundation for those who will go on to conduct research in any discipline. In addition, it can provide students with the skills needed to be more knowledgeable consumers of research. In the author's experience nearly all undergraduate students and many masters level students need assistance in order to move beyond reading the beginning and end of research articles while skipping the analytical portion of published work. The task need not fall only on those faculty teaching a research methods, statistics, or psychometrics class. Many faculty, across diverse disciplines, assign article reviews in their classes. Including in the assignment guidelines, critique of the analytical section of articles is an important step in improving the quantitative literacy of all graduates. If each faculty member selects for inclusion even a few design and/or analysis issues for students to consider, collectively, students are likely to acquire greater breadth in their understanding of research design and analysis matters. Instilling in students a healthy skepticism for published work along with the skills to detect problems could be of considerable value.

Faculty working with masters or doctoral students can then add considerable depth to a wide range of research topics and assignments which can include activities designed to prepare graduates to serve as journal peer reviewers (Zhu, 2014). Assigning 'Chaos in the Brickyard' or comparable pieces for reading and including reference to predatory publication practices would also complement efforts to enhance students' skills and knowledge with respect to research. This might not directly address the problems noted with respect to the generation of weak research but it will help students navigate the 'chaos in the brickyard'.

Turning to the service component of faculty responsibilities, there are a number of ways to promote quality research. Those involved in the retention, tenure and promotion review (RTP) process can help by engaging their colleagues in discussions that favor quality research over the simple quantity of publications. Without combating the publish or perish culture, the balance in expectations can still be shifted to the point where less is more. Since faculty are the ones sitting on RTP committees they can have a direct impact on keeping expectations with regard to quantity manageable. When connected to the efforts of faculty involved in shared governance (e.g., campus senates), university policies can be refined to value quality over quantity. This provides a framework for faculty and administrators reviewing tenure-track faculty and makes expectations clear to those being reviewed.

At first glance, it might not be clear how this would impact the 'chaos in the brickyard'. The connection is that faculty research is often done in the context of publish or perish expectations. This could lead to research choices driven by a need to quickly finish multiple projects which potentially floods the brickyard with small-scale unrelated findings based on data from few subjects. Reigning in a quantity-focused culture benefits everyone. We will reach a point of diminishing returns and more 'chaos in the brickyard' if the pressure to publish results in potentially weak research distributed through publishers with poor or nonexistent standards.

For those whose service takes the form of reviewing manuscripts for publication or presentation, their responsibilities serve a critical function in keeping the brickyard supplied with quality research. Requiring additional information of authors when needed including checks of distributional assumptions, p values, adjustments to alpha when multiple inferential tests are done, effect size(s), and practical significance will strengthen the end product prior to publication. When analyses are not familiar to a reviewer, asking the editor to solicit review of the analytical work could prove essential and result in important changes that otherwise might not have been made. Papers need not be rejected when lacking in one or more respects. Rather, modifications can be requested prior to accepting a manuscript. Such efforts benefit both reviewers and authors while strengthening the credibility of the journal's publications. The potential to change the proportion of strong vs. weak bricks that drive the construction of edifices is significant.

On a related note, journal editors have a gatekeeper role that impacts the quality of published work. Beyond the responsibilities of a reviewer, an editor in concert with their editorial board can explicitly establish the basic requirements for quantitative research and host/sponsor webinars or conference meetings for reviewers and authors to reinforce good practice. Additional recommendations have included publishing clear evaluation standards, clarifying roles among editors and reviewers, protecting the time commitment of editors and reviewers, and improving reviewer recognition (Knudson, Morrow & Thomas, 2014).

Faculty conducting research, independently or in collaboration with others, are in an excellent position to diminish the 'chaos in the brickyard'. A focus on good quality work from the design of their research and review of existing literature through the implementation of a project, data analysis, and write up of the findings will benefit the entire community of scholars as well as those who base decisions and actions on published findings.

Bartlett's (2013) example of the proliferation of flawed work makes clear that good quality research begins with understanding and questioning published work. We are cautioned early on in life to look before we leap. In the research domain it is important to probe other authors' work before attempting to build upon it. Otherwise, we run the risk of perpetuating weak ideas and leading others to pursue a misguided line of research.

With regard to the analysis of data, to the extent possible, each researcher should have enough of an understanding of basic descriptive and inferential statistics to insure that appropriate analyses are conducted; even when the person actually doing the analysis may be a paid consultant. The principal researcher should be the one guiding the research design and analysis of their data to insure that appropriate analyses are done to address well designed research questions. Once the data are analyzed and outcomes critically examined, findings regarding statistical significance (including p values) should be accompanied by measures of practical significance, power (post-hoc), and effect size. One final point with regard to the write up of a manuscript is that keyword selection should be taken seriously. Careful consideration of what descriptors others will need to find relevant publications is important.

**Conclusions/Implications**

The suggestions advanced here are certainly not a comprehensive list of all that can be done. Rather, they are meant to provide a catalyst for discussion and action on these and other ideas students, faculty, administrators, and editors might have. Members of the academic community, across all disciplines, can help by taking their own manageable share of responsibility for bringing order to the 'chaos in the brickyard'. The chaos did not emerge overnight; yet collectively if each person takes one small portion of the task in hand we can substantively change the landscape. Research of good quality provides us with information that advances our understanding of important issues in a sound and incremental manner.

References

Achterberg, C. L., Novak, J. D., & Gillespie, A. H. (1985). Theory-driven research as a means to improve nutrition education. *Journal of Nutrition Education*, 17(5), 179-184.

Altman, D. G, (2012). Building a metaphor: Another brick in the wall? *British Medical Journal*, *345*. doi: 10.1136/bmj.e8302. Retrieved from http://www.nih.gov/

Bartlett, T. (2013, August 5). The magic ratio that wasn't. *Chronicle of Higher Education.* Retrieved from http://chronicle.com

Bauerlein, M., Gad-el-Hak, M., Grody, W., McKelvey, B., & Trimble, S. W. (2010, June 13). We must stop the avalanche of low-quality research. *Chronicle of Higher Education.* Retrieved from http://chronicle.com

Beall, J. (2013). Beall's List of Predatory Publishers 2013. *Scholarly Open Access.* Retrieved from http://scholarlyoa.com/2012/12/06/bealls-list-of-predatory-publishers-2013/

Bower, B. (2013). Bias seen in behavioral studies. *Science News*, *184*(7), 10.

Chwe, M. S. (2014). Scientific pride and prejudice. *New York Times.* Retrieved from http://www.nytimes.com

Editorial commentary (2007). Dealing with biased reporting of the available evidence. *The James Lind Library.* Retrieved from www.jameslindlibrary.org

Eisenhardt, K. M. (1989). Building theories from case study research. *Academy of Management Review*, 14(4), 532-550.

Faul, F., Erdfelder, E., Lang, A. G., & Buchner, A. (2007). G*Power 3: A flexible statistical power analysis program for the social, behavioral, and biomedical sciences. *Behavior Research Methods*, *39*, 175-191.

Fischman, M. G. (2011). 'Chaos in the brickyard' revisited: What if Forscher were a butcher? *Research Quarterly for Exercise and Sport*, *82*(1), iii–iv.

Forscher, B. K. (1963). Chaos in the Brickyard. *Science*, *142*(3590), 339. doi: 10.1126/science.142.3590.339

Franks, B. D., & Huck, S. W. (1986). Why does everyone use the .05 significance Level? *Research Quarterly for Exercise and Sport*, *57*(3), 245-249.

Ioannidis, J. P. A (2005). Why most published research findings are false. *PLOS Medicine.* Retrieved from http://www.plosmedicine.org

Knudson, D. (2009). Significant and meaningful effects in sports biomechanics research. *Sports Biomechanics*, *8*, 96-104.

Knudson, D. V., Morrow J. R. & Thomas, J. R. (2014). Advancing kinesiology through improved peer review. *Research Quarterly for Exercise and Sport*, *85*(2), 127-135, doi: 10.1080/02701367.2014.898117

Kolata, G. (2013, April 7). For scientists, an exploding world of pseudo-academia. *New York Times.* Retrieved from http://www.nytimes.com

Lamb, E. (2013). How should we we write about statistics in public? *Scientific American*. Retrieved from http://www.scientificamerican.com/

Marteniuk, R. G., & Bertram, C. P. (1999). On achieving strong inference in prehension research. *Motor Control*, *3*, 272-275.

Myers, N.D., Ahn, S., & Jin, Y. (2011). Sample size and power estimates for a confirmatory factor analytic model in exercise and sport: A Monte Carlo approach. *Research Quarterly for Exercise and Sport*, *82*(3), 412-423.

Pigott, T. D., Valentine, J. C., Polanin, J. R., Williams, R. T., & Canada, D. D. (2013). Outcome-reporting bias in Education research. *Educational Researcher*, *42*(8), 424-432.

Seife, C. (2011). Context is everything-more about the dark arts of mathematical deception *Lecture at Google's New York Office*, Google, New York. http://www.youtube.com/watch?feature=player_embedded&v=qiQwZ6inbOM. Retrieved from http://ducknetweb.blogspot.com/.

Seife, C. (2014). What scientific idea is ready for retirement? Statistical Significance. *Edge.* Retrieved from http://www.edge.org/response-detail/25414

Silver, N. (2012). FiveThirtyEight's 2012 Forecast. *FiveThirtyEight Blog*. Retrieved from http://fivethirtyeight.blogs.nytimes.com/

Taleb, N. N. (2014). What scientific idea is ready for retirement? Standard Deviation. *Edge*. Retrieved from http://www.edge.org/response-detail/25401

Thomas, J. R., Nelson, J. K., & Silverman, S. J. (2010). Research methods in physical activity (6th edition). *Experimental and quasi-experimental research* (pp. 329-337). Champaign, ILL: Human Kinetics.

Thomas, J. R., Nelson, J. K., & Thomas, K. T. (1999). A generalized rank-order method for nonparametric analysis of data from exercise science: A tutorial. *Research Quarterly for Exercise and Sport*, *70*(1), 11-23.

Tromovitch, P. (2012). Statistical reporting with Philip's sextuple and extended sextuple: A simple method for easy communication of findings. *Journal of Research Practice*, *8*(1). Retrieved from http://jrp.icaap.org

Vacha-haase, T., Ness, C., Nilsson, J., & Reetz, R., (1999). Practices regarding reporting of reliability coefficients: A review of three journals. *The Journal of Experimental Education*. *67*(4), 335-341.

Vaisrub, N. (1991). Remembrance of things simple. *Chance: New Directions for Statistics and Computing, 4*(1), 52.

Walshe, K. (2007). Understanding what works - and why - in quality improvement: the need for theory-driven evaluation. *International Journal for Quality in Health Care*, 19(2), 57-59.

Weed, M. (2006). Research synthesis in the sport & exercise sciences: Introduction to the collection. *European Journal of Sport Science*, *6*(2), 93-95.

Weed, M. (2005). Research synthesis in sport management: Dealing with ''Chaos in the Brickyard''. *European Journal of Sport Science*, *5*(1), 77-90.

Weins, C. (2014). New truths that only one can see. *New York Times*. Retrieved from http://www.nytimes.com

Zhu, W. (2014). Together, we move science forward. *Research Quarterly for Exercise and Sport*, *85*(2), 125-126. doi: 10.1080/02701367.2014.904165